\documentclass[preprint,showpacs,preprintnumbers,amsmath,amssymb,nofootinbib]{revtex4}
\usepackage{booktabs}
\usepackage{mathrsfs}
\usepackage{epsfig}
\usepackage{graphicx}% Include figure files
\usepackage{dcolumn}% Align table columns on decimal point
\usepackage{bm}% bold math
\usepackage{amsmath}
\usepackage{slashed}
\usepackage{multirow}
\usepackage{color}

\def\bit{\begin{itemize}}
\def\eit{\end{itemize}}
\def\ben{\begin{enumerate}}
\def\een{\end{enumerate}}
\def\bed{\begin{description}}
\def\eed{\end{description}}

\def\k{\kappa}
\def\l{\lambda}

\def\cmg{\, {\rm cm^2/g} }
\def\half{\frac{1}{2}\,}
\def\third{\frac{1}{3}\,}

\def\lsim{\raise0.3ex\hbox{$<$\kern-0.75em\raise-1.1ex\hbox{$\sim$}}}
\def\gsim{\raise0.3ex\hbox{$>$\kern-0.75em\raise-1.1ex\hbox{$\sim$}}}

\let\jnfont=\rm
\def\NPB#1,{{\jnfont Nucl.\ Phys.\ B }{\bf #1},}
\def\PLB#1,{{\jnfont Phys.\ Lett.\ B }{\bf #1},}
\def\EPJC#1,{{\jnfont Eur.\ Phys.\ Jour.\ C }{\bf #1},}
\def\PRD#1,{{\jnfont Phys.\ Rev.\ D }{\bf #1},}
\def\PRL#1,{{\jnfont Phys.\ Rev.\ Lett.\ }{\bf #1},}
\def\MPLA#1,{{\jnfont Mod.\ Phys.\ Lett.\ A }{\bf #1},}
\def\JPG#1,{{\jnfont J.\ Phys.\ G}{\bf #1},}
\def\CTP#1,{{\jnfont Commun.\ Theor.\ Phys.\ }{\bf #1},}
\def\JHEP#1,{{\jnfont JHEP \ }{\bf #1},}
\def\NPPS#1,{{\jnfont Nucl.\ Phys.\ Proc.\ Suppl.\ }{\bf #1},}

\def\beq{\begin{equation}}
\def\eeq{\end{equation}}
\def\bea{\begin{eqnarray}}
\def\eea{\end{eqnarray}}
\newcommand{\ba}{\begin{array}}
\newcommand{\ea}{\end{array}}

\begin{document}

\title{PandaX limits on light dark matter with light mediator in the singlet extension of MSSM}

\author{Wenyu Wang$^1$}
\author{Jiajun Wu$^1$}
\author{Zhaohua Xiong$^1$}
\author{Jun Zhao$^{2,3}$}

\affiliation{$^1$ Institute of Theoretical Physics, College of Applied Science, 
Beijing University of Technology, Beijing 100124, China\\
$^2$ Key Laboratory of Theoretical Physics, Institute of Theoretical Physics, 
Chinese Academy of Sciences, Beijing 100190, China\\
$^3$ School of Physical Sciences, University of Chinese Academy of Sciences, 
Beijing 100049, China 
\vspace{1cm} }

\begin{abstract}
Using the latest PandaX limits on light dark matter (DM) with light mediator,
we check the implication on the parameter space of the general singlet extension of MSSM
(without $Z_3$ symmetry),
which can have a sizable DM self-interaction to solve the small-scale structure problem.    
We find that the PandaX limits can stringently constrain such a paramter space, depending on 
the coupling $\lambda$ between the singlet and doublet Higgs fields.
For the singlet extension of MSSM with $Z_3$ symmetry, the so-called NMSSM, we also
demonstrate the PandaX constraints on its parameter space which gives a light DM 
with correct relic density but without sufficient self-interaction to solve the 
small-scale structure problem. We find that in this NMSSM the GeV dark matter  
with a sub-GeV mediator has been stringently constrained.  
\end{abstract}

\maketitle

\section{Introduction}
The large-scale structure of the universe ($> 1$ Mpc) can be successfully described by the 
$\Lambda$CDM (Lambda cold dark matter) cosmological model. However, the observation of 
small-scale structures~\cite{Bringmann:2013vra}, such as the Milky way and dwarf galaxies, 
seems to have tension with the simulations of collisionless cold dark matter. 
This dilemma is usually known as three problems: ~\textit{missing satellites} \cite{Klypin:1999uc},
~\textit{cusp vs ~core} \cite{deNaray:2011hy} and ~\textit{too big to fail} 
~\cite{BoylanKolchin:2011de}. 
A possible common solution to these issues is that the cold dark matter may have 
nontrivial self-interactions, namely the self-interacting dark matter (SIDM).
Such SIDM can be realized in the DM models with a light 
mediator~\cite{Feng:2008mu,Foot:2014uba,Blennow:2016gde}. 

Note that the SIDM models with a light force carrier ($\lsim$ 100 MeV) can have 
a non-trivial velocity-dependent scattering cross section and may explain the
small-scale structure problems ~\cite{Tulin:2012wi,Ko:2014nha}, which have been 
widely studied in the past few years. However, if the interaction between DM particle 
and the target nuclei is induced by a light mediator, the scattering cross section 
will be enhanced at low momentum transfer~\cite{Foot:2003iv,Li:2014vza,DelNobile:2015uua,Kahlhoefer:2017ddj}. The current DM direct detection experiments have reached impressive sensitivities 
and are approaching the irreducible background neutrino floor. The null results produced 
strong constraints for various DM models. In particular, the PandaX-II collaboration has 
recently searched for the nuclear recoil signals of DM with light mediators~\cite{Ren:2018gyx}. 
Using data collected in 2016 and 2017 runs, they set strong upper limits on the DM-nucleon 
cross section for different mediator masses. On the other hand, if the light mediator 
mainly decays into the SM particles, such as $\gamma\gamma$ and $e^+e^-$, the injection 
of sizable energy densities into the visible sector thermal bath after the light elements 
may spoil the success of the big bang nucleosynthesis (BBN). This will produce a lower limit 
on the couplings between mediator and the SM particles and then may lead to a tension with 
the direct detections.

In this work, we investigate the implication of the PandaX limits on the parameter space of the 
singlet extension of MSSM with or without $Z_3$ symmetry ~\cite{fayet,NMSSM,Kaminska:2014wia}.
Such models can alleviate the little hierarchy problem by pushing
up the SM-like Higgs boson mass to 125 GeV without heavy top-squarks \cite{Cao:2012fz}
and the model with $Z_3$ symmetry, the so-called NMSSM,  
can solve the notorious $\mu$-problem in the MSSM~\cite{muew} by dynamically 
generating the SUSY-preserving $\mu$-term. 
For a rather small $\lambda$ (the coupling between the singlet and doublet Higgs fields), 
the singlet sector can be almost decoupled from the electroweak
symmetry breaking sector and becomes a hidden sector of the model. 
The singlino-like DM will dominantly annihilates into the SM particles through 
the $s$-channel light singlet-like Higgs bosons to produce the correct 
DM relic density~\cite{NMSSM2,Hooper:2009gm,Wang:2009rj}. 
Note that the presence of light mediator can lead to long-range attractive 
forces between DM particles and enhance the DM annihilation cross section via 
the Sommerfeld effect at low temperature \cite{Wang:2014kja}. 
The previous study \cite{Wang:2014kja} showed the general singlet extension of MSSM without 
$Z_3$ symmetry (hereafter is called GNMSSM) can have a sizable DM self-interaction to 
solve the small-scale structure problem,
while the NMSSM can give a light DM 
with correct relic density but without sufficient self-interaction to solve the 
small-scale structure problem. 

The paper is organized as follows. In Sec. \ref{sec2} we focus on the NMSSM and  
examine the limits on the parameter space which has a light dark matter with light 
mediator. 
In Sec. \ref{sec3} we first show the parameter space of the GNMSSM which can 
solve the small-scale structure problem, and then check the PandaX limits on the
parameter space.
Our conclusion is given in  Sec. \ref{sec4}.

\section{Constraints on the NMSSM}
\label{sec2}
Since the NMSSM with $Z_3$ symmetry can give a light DM 
with correct relic density but without sufficient self-interaction to solve the 
small-scale structure problem \cite{Wang:2014kja}, in the following we do not require
it to solve the small-scale structure problem.  

In the NMSSM the superpotential of the doublet and singlet Higgs fields is given by
\begin{eqnarray}
\lambda\hat{S}\hat{H_u}\cdot\hat{H_d}+\frac{\kappa}{3}\hat{S}^3 \, ,\label{sp-nm}
\end{eqnarray}
where $\hat{H}_u$ and $\hat{H}_d$ are the Higgs doublet superfields, $\hat{S}$ is the
singlet superfield, and $\lambda$ and $\kappa$ are dimensionless parameters.
Note that the $Z_3$ symmetry is imposed in the superpotential to forbid
other terms of the singlet. The corresponding soft SUSY breaking terms are given by
\begin{eqnarray}\label{soft-term-nm}
A_\lambda \lambda S H_u\cdot H_d+\frac{A_\kappa}{3}\kappa S^3 +h.c.\, .
\end{eqnarray}
Then we can get three CP-even Higgs bosons (denoted as
$h_{1,2,3}$), two  CP-odd Higgs bosons (denoted as  $a_{1,2}$) and a pair of 
charged Higgs bosons. 
From Eq. (\ref{sp-nm}), we can also see that the interactions between the singlet
and the SM sector are controlled by the parameter $\l$. 
The constraints on the singlet-like Higgs bosons and siglino-like DM 
from both collider experiments and dark matter detections 
can be satisfied if $\l$ is sufficiently small.
Since the spectrum of the NMSSM has been widely studied in the literature \cite{NMSSM},
here we concentrate on the dark singlet sector.

In the NMSSM, the dark matter can have three components: gaugino, higgsino and singlino.
Assuming the gaugino unification relation $M_2/M_1\approx 2$,
we have three dark matter scenarios:
\begin{itemize}
\item Bino-dominant dark matter. As shown in \cite{Cao:2010ph}, under current collider
and DM relic density constraints, the SI cross section can exclude a large part
of parameter space, especially, if such bino-like DM contains a sizable 
higgsino component, its scattering cross section with the nucleon may be large and 
thus subject to stringent constraints from the direct detection limits \cite{Abdughani:2017dqs}.
\item Higgsino-dominant dark matter. As pointed in \cite{Wang:2013rba}, the higgsino-dominant DM
candidate around 1.1 TeV can satisfy all the constraints, including the relic density
and current DM direct detections.  
\item Singlino-dominant dark matter. In order to explain the observation of
CoGeNT \cite{Aalseth:2010vx}, the analysis in \cite{Draper:2010ew}
showed that in the Peccei-Quinn limit there can
exist three light singlet-like particles (0.1-10 GeV): a scalar,
a pseudoscalar and a singlino-like dark matter.
For a certain parameter window, through annihilation into the light pseudoscalar
the singlino dark matter can give the correct relic
density, and through exchanging a light scalar in scattering off the nucleon
a large cross section suggested by CoGeNT and DAMA/LIBRA \cite{Bernabei:2010mq}
can be attained.
\end{itemize}
%%%%%%%%%%%%%%%%%%%%%%%%%%%%%%%%%%%%%%%%%
Since the first two scenarios are the same as in the MSSM, 
we focus on the third scenario which is called the dark light Higgs (DLH) scenario.
As studied in \cite{Draper:2010ew},
besides explaining the observation of CoGeNT and DAMA/LIBRA, the parameter space
is also consistent with the experimental constraints form LEP, the Tevatron, $\Upsilon$
and flavor physics. In fact, the DLH scenario is the only possibility
for realizing the self-interaction in the NMSSM. 
However, as studied in \cite{Wang:2014kja}, the direct detection limits
give stringent constraints on the couplings between dark matter and SM particles,
inducing a much small $\l$. The will make $\sigma_\chi/m_\chi$ too small
to explain the small cosmological scale simulations.

In the following,  we check the PandaX constraints on the DLH
scenario which can have light singlino-like DM and light singlet-like Higgs 
bosons as the mediators. 
The light singlino-like DM particles mainly annihilate to SM particles 
through the resonance of the singlet-like  pseudo-scalar $a_1$. 
The DM scatters off the nucleon with the light CP-even
singlet-like Higgs boson $h_1$ as the mediator, and the scattering cross section
is subject to the PandaX limits.
Following Ref. \cite{Draper:2010ew}, in order to approach the PQ symmetry limit, we
define the parameter $\varepsilon\equiv {\lambda \mu}/{m_Z}$,
$\varepsilon' \equiv {A_\lambda}/{\mu\tan\beta} -1$.
We perform a random scan\footnote{We did not use the machine learning scan 
\cite{Ren:2017ymm} 
because the parameter space is not too large.} 
over the parameter space: $2\leq\tan\beta\leq 50$, $0.05\leq\lambda\leq 0.2$,
$0.0005\leq\kappa\leq0.05$, $-0.1\leq\varepsilon \leq 0.1$,
$\varepsilon \sim \varepsilon'$, $|A_\k|<500{\rm GeV}$, and $|\mu|<1000{\rm GeV}$.
The sfermion sector parameters are set at 6 TeV so that we can get a 125 GeV SM
Higgs boson easily. 
In our scan we use the package NMSSMTools \cite{nmssmtools} to obtain the parameter space with
a singlet-like Higg boson $h_1$ lighter than 2 GeV. 
We require the DM thermal relic density in the 2$\sigma$ range of the Planck value \cite{planck} 
and the mass of the SM-like Higgs $h_2$ in the range of 123-127 GeV.

%%%%%%Fig.1%%%%%%%%%%%%%%%%%%%%%%%%%%%%%%%%%%%%
\begin{figure}
\begin{center}\scalebox{0.8}{\epsfig{file=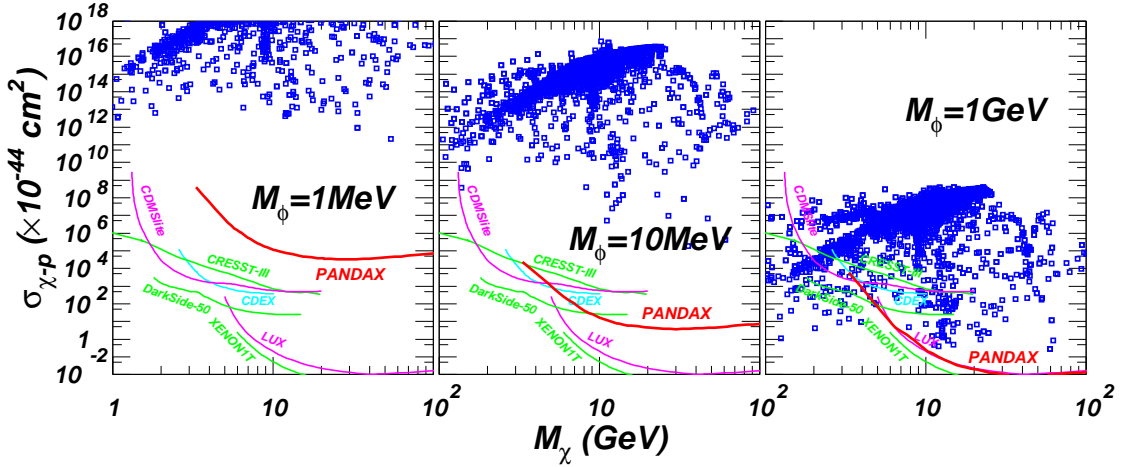}}\end{center}
\vspace{-.5cm}
\caption{Scatter plots showing the spin-independent cross section between dark matter 
and nucleon with different mediator masses in the DLH scenario in the NMSSM with $Z_3$ symmetry.
The light mediator $\phi$ is the lightest CP-even Higgs boson $h_1$ which is singlet 
dominant. All the points satisfy the constraints of the DM relic density and the SM like Higgs
boson $h_2$ in the range of 123-127 GeV.
The detection limits of PandaX \cite{Ren:2018gyx} on light mediator are shown as the red 
lines.
The detection limits of CDEX-10 \cite{cdex}, CDMSlite Run 2 SI\cite{cdmslite},
CRESST-III \cite{cresst}, LUX Combined\cite{lux-2017}, and XENON1T \cite{xenon1t}  are also
shown.}
\label{fig1}
\end{figure}
%%%%%%%%%%%%%%%%%%%%%%%%%%%%%%%%%%%%%%%%%%%%%%%

The results are shown in Fig. \ref{fig1}, in which we set the mass of $h_1$
to 1 MeV, 10 MeV and 1 GeV, respectively. 
In each panel we show the detected limits of PandaX under the corresponding
mediator masses. For comparison, we also show the detection limits of
CDEX-10 \cite{cdex}, CDMSlite Run 2 SI\cite{cdmslite},
CRESST-III \cite{cresst}, LUX Combined\cite{lux-2017}, and XENON1T \cite{xenon1t}.
We can see that the SI cross section is enhanced greatly as
the mass of the light mediator decreases. This can be understood from 
the cross section \cite{Draper:2010ew}
\begin{eqnarray}
\sigma_{\rm SI} \approx
\frac{\left [\left(\frac{\varepsilon}{0.04} \right)+ 0.46 \left (\frac{\lambda}{0.1} \right)
\left(\frac{\
v}{\mu}\right) \right]^2 \left(\frac{y_{h_1\chi\chi}}{0.003}\right)^2 10^{-40}
\mathrm{cm}^2}{ \left(\
\frac{m_{h_1}}{1 \mathrm{GeV}}\right)^4 },
\end{eqnarray}
where $y_{h_1\chi\chi}$ is the coupling strength of the Higgs boson $h_1$ and dark matter.
This relation implies that the cross section increases
as the fourth power of the inverse mass of light mediator.
If the mass of the mediator is at the order of MeV, a lot of samples will
exceed the detection limits and thus the DLH scenario of the NMSSM
is severely constrained  by PandaX and other experiments (except for the region outside
the detection sensitivity, which has the dark matter lighter than 3.5 GeV for the
PandaX results). Such stringent constraints are from the correlation between
the dark matter relic density and the dark matter-nucleon scattering cross section:
a proper relic density implies a non-negligible coupling $a_1f\bar f$ and 
the parameter $\l$ can not be too small, and so the SI cross section 
will be enhanced by the mediator mass greatly, as shown in the above equation.

From Fig. \ref{fig1} we see that the dark matter with a mediator at MeV order is
also excluded by other direct detections.
Only a small  parameter space with a mediator mass around GeV and a dark matter 
mass is around several GeV can survive all the direct detection limits.

\section{Constraints on the GNMSSM}\label{sec3}

Compared with the NMSSM with $Z_3$ symmetry,  the
general singlet extension of MSSM (GNMSSM) has a larger 
parameter space, in which the $Z_3$ discrete symmetry is not imposed and
the $\mu$ term can exist in the superpotential, together with
the $\l S H_u \cdot H_d$ term (several other terms
of the singlet superfield can also exist in  the superpotential).
Consequently, the dark Higgs sector (including a singlino-dominant dark matter)
can be easily realized in the GNMSSM, which does not need the condition
 $\k \ll \l$. This means that a singlino-dominant dark matter can be
obtained with a sizable $\k$ and in this case the coupling
$h_1\chi\chi$ in dark matter self-interaction,
which is proportional to $\k$, can be large.

As pointed in \cite{Tulin:2013teo},  in order to solve the
small-scale simulation anomalies, the self-interaction between the dark matter is needed. 
In the non-relativistic limit,
the scattering between dark matter can be described by quantum mechanics.
The most recent simulations have shown that the ratio of between cross section
$\sigma$ and the mass of the dark matter $m_\chi$ on
on dwarf scales (the characteristic velocity is 10 km/s)
should be $\sigma/m_\chi \sim 0.1 - 10 \, \cmg$ to solve the
\textit{core-vs-cusp} and \textit{too-big-to-fail} problems,
while the Milky Way (the characteristic velocity is 200 km/s)
and cluster scales (the characteristic velocity is 1000 km/s) require
$\sigma/m_\chi \sim 0.1 - 1 \, \cmg$.  It appears that all the data may be
accounted for with a constant scattering cross section around
$\sigma/m_\chi \sim 0.5 \, \cmg$.
The numerical input for the simulation of
small scales is the differential cross section
$d \sigma/ d \Omega$, which is a function of the dark matter relative velocity $v$.
Then the viscosity (or conductivity) cross section $\sigma_V$ ~\cite{krstic:1999} can
be defined as
\beq
\sigma_V =  \int d \Omega \, \sin^2 \theta \, \frac{d\sigma}{d\Omega} \, ,  \label{sigmaT}
\eeq
where the weight of $\sin^2 \theta$ is needed since both forward
and backward scattering amplitudes will diverge.
Such singularities correspond to the poles in
the $t$- and $u$-channel diagrams for the identical dark matter candidate.
Within the resonance region, $\sigma_V$ must be computed by solving the Schr\"{o}dinger
equation. Detail of the solution can be found in  \cite{Wang:2014kja}.
As there are   symmetric ($\sigma_{VS}$) and antisymmetric ($\sigma_{VA}$)
cross section for Majorana fermion dark matter  models.
In our following analysis, we assume that the dark matter
 scatters randomly. Thus the average cross  section will be
\bea
\sigma_{V} =\frac{1}{4}\sigma_{VS}+\frac{3}{4} \sigma_{VA}\; .\label{sigmaVa;;}
\eea

Now we turn to the dark Higgs sector of the GNMSSM for the explanation of the small cosmological
structure problem. The renormalizable superpotential for the singlet is given by 
\bea
W = \eta \widehat{S} +\half \mu_s \widehat{S}^2 + \frac{1}{3} \k \widehat{S}^3 \ ,\label{sp-gm}
\eea
and the soft SUSY breaking terms take the form
 \bea
 -{\cal L}_\mathrm{soft} &=&  m_s^2 | S |^2
 + \left( C_\eta \eta S +\half B_s \mu_s S^2 + \third \k A_\kappa\ S^3 + \mathrm{h.c.}
 \right)\,. \label{soft-gm}
\eea
where $\eta, \mu_s, m_s^2, C_\eta, B_s$ are the additional input parameters
in the GNMSSM, compared with the NMSSM.  
After the scalar component gets a VEV,  we can get one CP-even Higgs $h$ and
one CP-odd Higgs $a$. The detail of spectrum and Feynman rules can be found in \cite{Wang:2014kja}.
Though this singlet sector is the dark sector, it can give a correct relic density
of dark matter and proper self-interaction.

%%%%%%Fig.2%%%%%%%%%%%%%%%%%%%%%%%%%%%%%%%%%%%%
\begin{figure}
\begin{center}\scalebox{0.4}{\epsfig{file=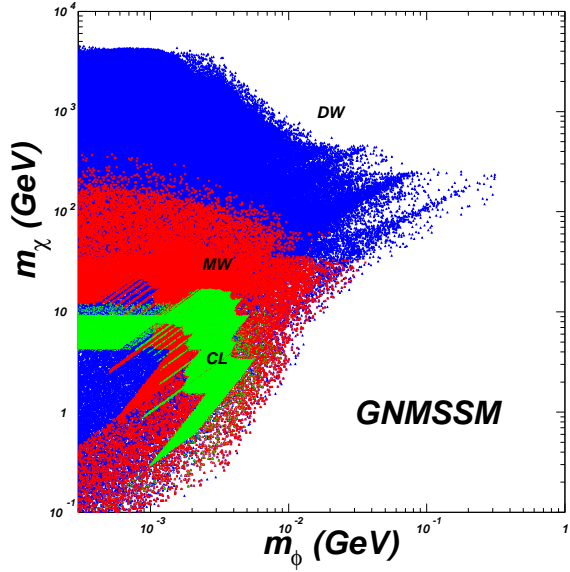}}\end{center}
\vspace{-.5cm}
\caption{The scatter plots of the GNMSSM satisfying the DM relic density, 
taken from \cite{Wang:2014kja}.
The blue points satisfy the simulation in the Dwarf scale,
while the red and green points satisfy the simulation in the Milky Way with
a characteristic velocity of 200 km/s and 1000 km/s, respectively.)}
\label{fig2}
\end{figure}
%%%%%%%%%%%%%%%%%%%%%%%%%%%%%%%%%%%%%%%%%%%%%%%

A detailed study of the GNMSSM in solving the small cosmological scale anomalies
is presented in \cite{Wang:2014kja}. Here we just demonstrate the parameter space 
in Fig. \ref{fig2}.
We can see that a part of the parameter space can satisfy simultaneously the requirements of
the dwarf, the Milky Way and galaxy cluster scales to solve the small scale anomalies.
In this part of the parameter space the masses of dark matter and mediator are quite 
restrained. 

For the connection between the electroweak sector and the dark light singlet sector,
the mixing angles between singlet and doublet fields depend on the off-diagonal elements 
divided by the differences between diagonal elements. For example, the mixing
between $H_d$ and $S$ is proportional to
\begin{eqnarray}
 \theta_{ds}\sim \frac{{\cal M}_{S,13}^2}{\left|{\cal M}_{S,11}^2
-{\cal M}_{S,33}^2\right|} \sim \l \times \left(\mbox{electro-weak variables}\right)
\end{eqnarray}
where ${\cal M}$ is the $3\times 3$ Higgs mass matrix \cite{fayet,NMSSM,Kaminska:2014wia}.
If the mediator $h_1$ is around several MeV, the input electroweak parameters
give  ${\cal M}_{S,13}^2$ and ${\cal M}_{S,33}^2$ much smaller than
the electroweak scale. Thus we can define the following two angles
for the mixing between singlet and doublet fields
\begin{eqnarray}
  \theta_{ds}=\l \alpha_d \cos\beta,~~\theta_{us}=\l \alpha_u \sin\beta,
\end{eqnarray}
where $\alpha_d$ and $\alpha_u$ are two new parameters for
the mixing angles. 
By such a parameterization, we can calculate the corresponding cross section.
For example, the coupling strength between singlet and up type quarks is given by
\begin{eqnarray}
  Y_q \l \alpha_u\sin\beta,
\end{eqnarray}
and for the down type quarks is given by 
\begin{eqnarray}
  Y_q \l \alpha_d\cos\beta.
\end{eqnarray}

%%%%%%Fig.3%%%%%%%%%%%%%%%%%%%%%%%%%%%%%%%%%%%%
\begin{figure}
\begin{center}\scalebox{0.6}{\epsfig{file=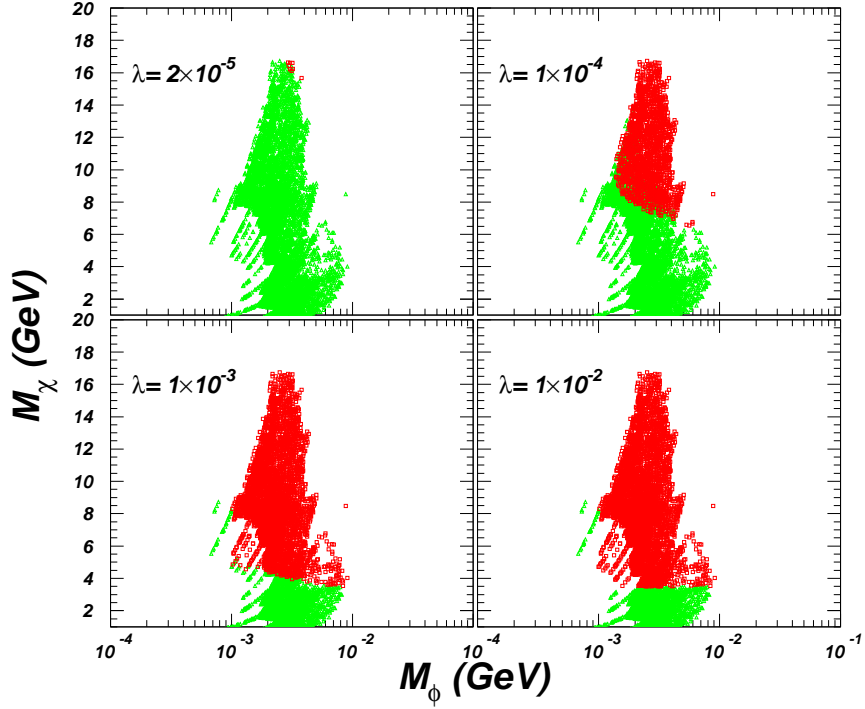}}\end{center}
\vspace{-.5cm} 
\caption{Scatter plots of the GNMSSM parameter space which can satisfy 
the dark matter relic density and solve  
the small cosmological structure problem, displayed on the plane of
the dark matter mass versus the mediator mass.   
The red points are excluded by the PandaX limits.}
\label{fig3}
\end{figure}
%%%%%%%%%%%%%%%%%%%%%%%%%%%%%%%%%%%%%%%%%%%%%%%
By the parameterizations above, we can calculate the spin-independent 
DM-nucleon scattering cross section \cite{Drees,Belanger}:
\begin{eqnarray}
\sigma^{SI} = \frac{4 m_r^2}{\pi} f_N^2
\end{eqnarray}
where
\begin{eqnarray}
  m_r = \frac{m_\chi m_N}{ m_\chi + m_N }
\end{eqnarray}
is the reduced dark matter mass, and $f_N$ is the effective coupling 
of the DM with nucleon.
Since the singlet sector is much lighter than the electroweak scale, we
can neglect the contribution of the squarks and the supersymmetric loop contribution. 
Thus $f_N$ can be written as
\begin{eqnarray}
    {f_N \over m_N} &=& \sum_{q=u,d,s} {f_{Tq}^N \over m_q}f_q
     + {2\over 27} f_{TG}^N \sum_{q=c,b,t} {f_q^{H} \over m_q},  \label{effcoup}
\end{eqnarray}
where $f_{Tq}^N$ denotes the fraction of the nucleon mass $m_N$ that is due to 
the light quark $q$, and
\begin{eqnarray}
 f_{TG}^N=\frac{2}{27}( 1 - f_{Tu}^N - f_{Td}^N - f_{Ts}^N )
\end{eqnarray}
is the heavy quark contribution to $m_N$,  which is induced via gluon exchange.
A detailed calculation of these parameters $f_{Tq}$ can be found in \cite{Drees}.
In our calculation, we use $\sigma_{\pi N}=64$ MeV and $\sigma_0 = 35$ MeV
to get the values of $f_{Tq}^N$.

Fig. \ref{fig3} shows the PandaX constraints on the GNMSSM parameter space in which
all the points satisfy the DM relic density and the
scattering cross section for the small cosmological structures.
Here we set $\tan\beta = 2$ and $\alpha_u=\alpha_d = 0.001$ for the demonstration.
We also checked the results of other different values of $\tan\beta$
and $\alpha_u, \alpha_d$, and found that the results are similar.
We can see that when $\l$ is less that of $10^{-5}$,
the dark sector can survive safely. As $\l$ increases,
the constraints become stringent.

Note that since there exist light singlet Higgs bosons and singlino-like DM, 
the SM-like Higgs will have additional decay channels 
$h_{SM}\to h_1 h_1$, $h_{SM}\to a_1 a_1$ and $h_{SM}\to \chi \chi$,
in which the first two channels are exotic decays and the last one is the invisible decay. 
Their branching ratios are determined by the coupling parameters $\l$ and $\k$.
These decays can give interesting phenomenology \cite{Wang:2016lvj}

We should note that the singlet CP-even Higgs boson can not be too dark,
it must decay before the start of the Big Bang Nucleosynthesis (BBN) ($\sim 1$ sec)
so its decay products will not affect BBN. But if the singlet Higgs couples to the
SM particles sizably, the direct detection rate of dark matter will be enhanced greatly
by the light mediator.
One way to solve such an obstacle is to add a right-handed
neutrino to the GNMSSM \cite{Kang:2016xrm}.
Another point we would like to adress is that the light dark matter in our analysis 
is around GeV scale. For sub-GeV ultra light dark matter, the dark matter particles can be
boosted by the cosmic rays and the detection sensitivity can be much enhanced \cite{sub-gev-dm}.
Also, for a heavy dark matter above TeV, recently the DAMPE collaboration reported 
the cosmic $e^+ + e^-$ flux excess \cite{Ambrosi:2017wek} 
which seems to favor a TeV-scale leptophilic dark matter \cite{leptophilic-dm-tev}.

\section{conclusion}\label{sec4}
Using the newest limits given by the PandaX collaboration 
on the zero-momentum dark matter-nucleon cross section,
we checked the implication on the supersymmetric
dark models, especially on the parameter space of light dark matter and mediator.
We first analyzed the spectrum of NMSSM with $Z_3$ symmetry and GNMSSM without $Z_3$ symmetry,
and found out a way to parameterize the connection between singlet sector
and the SM sector. Then we examined the parameter space of the two models under
the limits of PandaX and other direct detections.
We obtained the following observations:
\begin{itemize}
\item The PandaX limits exclude the case of dark matter above 
3.5 GeV. The left space are excluded by the requirement of the self-interaction
which gives a stringent constraints on the mass of mediator.
The reason is the correlation between
the dark matter relic density and the dark matter nucleon cross section.
Thus, the NMSSM with $Z_3$ symmetry is excluded by the PandaX and the requirement
of explanation for the small structure problems.
\item  It is easy to realize self-interaction in the GNMSSM,
in which the singlet sector can be a dark sector. In the dark sector,
the correct relic density and a proper self-interaction can be obtained. 
Compared to the
simple one-mediator model, the supersymmetric model can have a larger parameter space,
and the mass of the dark matter and mediator can be relaxed.
\item By our parameterization for the connection between singlet sector and the SM sector,
we find that PandaX results can give a constraint on the coupling strength between
the two sector. Only a very small $\l$ is allowed.
\end{itemize}
In all, the PandaX limits give us a very good test for the self-interaction dark matter
models. More precision measurements of the light mediator and dark matter, together with
the self-interaction physics need to be further studied.

\section*{Acknowledgments}
This work was supported by the Natural Science Foundation of China under grant number 11775012.

\end{document}